# Extreme flow simulations reveal skeletal adaptations of deep-sea sponges.

Giacomo Falcucci [1,2]✉, Giorgio Amati [3], Pierluigi Fanelli [4], Vesselin K. Krastev [1], Giovanni Polverino [5], Maurizio Porfiri [6,7,8,11] & Sauro Succi [2,9,10,11]

1: Department of Enterprise Engineering "Mario Lucertini", University of Rome "Tor Vergata", Rome, Italy.

2: Department of Physics, Harvard University, Cambridge, MA, USA.

3: High Performance Computing Department, CINECA Rome Section, Rome, Italy.

4: DEIM, School of Engineering, University of Tuscia, Viterbo, Italy.

5: Centre for Evolutionary Biology, School of Biological Sciences, University of Western Australia, Perth, Western Australia, Australia.

6: Department of Biomedical Engineering, Tandon School of Engineering, New York University, New York, NY, USA.

7: Department of Mechanical and Aerospace Engineering, Tandon School of Engineering, New York University, New York, NY, USA.

8: Center for Urban Science and Progress, Tandon School of Engineering, New York University, New York, NY, USA.

9: Italian Institute of Technology, Center for Life Nano- and Neuro-Science, Rome, Italy.

10: National Research Council of Italy – Institute for Applied Computing (IAC), Rome, Italy.

11: These authors contributed equally: Maurizio Porfiri, Sauro Succi.

✉e-mail: giacomo.falcucci@uniroma2.it

**Abstract**

Since its discovery[1,2], the deep-sea glass sponge *Euplectella aspergillum* has attracted interest in its mechanical properties and beauty. Its skeletal system is composed of amorphous hydrated silica and is arranged in a highly regular and hierarchical cylindrical lattice that begets exceptional flexibility and resilience to damage[3–6]. Structural analyses dominate the literature, but hydrodynamic fields that surround and penetrate the sponge have remained largely unexplored. Here we address an unanswered question: whether, besides improving its mechanical properties, the skeletal motifs of *E. aspergillum* underlie the optimization of the flow physics within and beyond its body cavity. We use extreme flow simulations based on the 'lattice Boltzmann' method[7], featuring over fifty billion grid points and spanning four spatial decades. These in silico experiments reproduce the hydrodynamic conditions on the deep-sea floor where *E. aspergillum* lives[8–10]. Our results indicate that the skeletal motifs reduce the overall hydrodynamic stress and support coherent internal recirculation patterns at low flow velocity. These patterns are arguably beneficial to the organism for selective filter feeding and sexual reproduction[11,12]. The present study reveals mechanisms of extraordinary adaptation to live in the abyss, paving the way towards further studies of this type at the intersection between fluid mechanics, organism biology and functional ecology.

Progress in computational science has disclosed unique insights across virtually all fields of science, including the emerging frontier between physics and biology[13,14]. The approach to exascale-class computational facilities[15] promises unprecedented scientific analyses and predictions, allowing the exploration of questions beyond the reach of experimental investigations. As an example, the study of the profound connection between structure and function lies at the core of modern biology, from cell cycles and metabolic analysis, all the way up to entire organs or the overall body of living creatures[16]. The availability of high-performance computational power is enabling the study of realistically complex biological models, including deep-sea organisms that are often inaccessible to in vivo experimentation. The main focus of the present work is the study of the fluid dynamic performance of the deep-sea glass sponge *E. aspergillum* in its actual living conditions, reproduced via in silico experiments on the high-performance Marconi100 system[17]. Taking advantage of the versatility and resolution provided by extreme numerical simulations, we analysed different flow regimes corresponding to the actual living conditions of *E. aspergillum* (shown in Figure 1a). Through our simulations, performed via the lattice Boltzmann method[7], we disentangled morphological characteristics in order to pin down the specific fluid dynamic role of each of the skeletal features. The deep-sea glass sponge *E. aspergillum* (also known as Venus' flower basket) belongs to the oldest group of sponges (Hexactinellida) documented in the phylum Porifera[18]: it dwells at the bottom of the oceans, especially in the Pacific and around Antarctica, at depths of 100−1,000 m, with no ambient sunlight[19,20]. The bottom of the body is anchored to the soft sediment of the sea floor and protrudes into the benthic boundary layer flow, thereby providing a substrate for suspension feeders. Interestingly, a breeding pair of shrimps (Spongicolidae family) that enter the cavity of the sponge through the small fenestrae on its surface will often be trapped in an eternally monogamous relationship as they grow in size—in Japan, they are called 偕老同穴 'Kairou Douketsu', which means 'together for eternity'[21,22]. *E. aspergillum* is characterized by exceptional structural properties that have sparked the attention of researchers since its discovery[1,23,24]. More specifically, the hierarchical arrangement of its skeletal system has proven to delay crack propagation and increase buckling strength[5,25], thereby approach- ing an optimal material distribution[6]. Despite the vast literature on the mechanical properties of *E. aspergillum*'s skeletal system, to the best of our knowledge, the study of the surrounding and internal hydrodynamics has not yet been performed to date.

We conducted extreme fluid dynamic simulations in order to assess the as-yet unknown hydrodynamic function of the different skeletal features of the sponge, under realistic living conditions, reproduced through different values of the water speed, and hence different Reynolds numbers. The Reynolds number is given by Re= $|\mathbf{u}|D/v$, where $\mathbf{u}$ is the water velocity, $D$ the sponge diameter at its top section (about 4 cm in our case[3]), and $v$ is the water kinematic viscosity ($v \approx 1.75 \times 10^{-6}$ m$^2$ s$^{-1}$ (ref. [10]), with potential variations of at most 15% due to modest changes in temperature and salinity in the abyss[26]; see also http://web.mit.edu/seawater). On the basis of the known, average distribution of water speed in proximity of the bottom of the ocean, which ranges between 0 and about 11 cm s$^{-1}$ in the first 30 cm over the sea bed[9], we analysed the flow regimes corresponding to Re = 100, 500, 1,000, 1,500 and 2,000. These simulations provide unique insights into the ecological adaptions of *E. aspergillum* to live in extreme environments[27], revealing the flow-driven raison d'être for the sponge's shape and its peculiar skeletal motifs.

For the complete model of *E. aspergillum* shown in Figure 1b, simulations at statistical steady state show a substantial reduction of the flow speed inside the body cavity of the sponge, where low-speed vortical flow patterns are formed (see Figure 1c). Surprisingly, such a quiescent region extends downstream of the sponge in contrast with the wake behind solid obstacles[28]. Only several diameters downstream do we observe the emergence of intermittent patterns[29,30]. The complexity of the flow inside and outside the body cavity is evidenced by means of the contours of helicity ($\mathcal{H} = \mathbf{u} \cdot \boldsymbol{\omega}$, where $\boldsymbol{\omega} = \nabla \times \mathbf{u}$ is the local flow vorticity, see Methods) and the streaklines coloured according to the flow speed (Figure 1c): the presence of the quiescent region extending several diameters downstream of the specimen is apparent from our in silico experiments (see Extended Data Figure 2 and Extended Data Figure 3 and Supplementary Information Video 1 for more details).

To shed light on such a peculiar fluid dynamic behaviour and to clarify its dependence on *E. aspergillum*'s skeletal motifs, we conducted a series of in silico morphological manipulations. We examined four simplified models of increasing complexity, ranging from a solid cylinder to a hollow cylindrical lattice with helical ridges, which proxies the real morphology of the sponge without the anchoring to the seafloor. More specifically, we studied two solid models (S1 and S2; a plain cylinder, S1, and a cylinder with the same helical ridge patterns that decorate *E. aspergillum*'s skeletal system, S2) and two porous models (P1 and P2, with fenestrae on their surface; a hollow cylindrical lattice, P1, and a hollow cylindrical lattice with helical ridges, P2). All of these geometries, reported in Figure 1b, are subjected to periodic boundary conditions at the top and bottom. For each simplified model, we simulated the corresponding hydrodynamic field, and studied the downstream intermittency that arises for Re ≥ 100. We monitored the evolution in time (*t*) of the fluid velocity **u**(*P*, *t*) at a probe *P* located at a distance 2.5*D* downstream of the model, *D*/2 from the symmetry axes along the flow direction, and *H*/2 from the bottom (*H* being the height of our computational domain). Figure 2a reports the time evolution of the three Cartesian components of the flow velocity $u_x(t)$, $u_y(t)$, $u_z(t)$ at the probe location *P*. Figure 2a refers to the final ~1% of the time evolution, well within the statistical steady state of the given fluid regime. Results in Figure 2a indicate that the two solid models are characterized by intense fluctuations of all three velocity components. By contrast, even at Re = 2,000, the porous models exhibit a substantial reduction of such fluctuations. Moreover, we found that even at high Reynolds numbers (Re ≥ 500), the presence of the fenestrae in the porous models abates the *z* component of the fluid velocity, leading to a quasi-2D flow in the wake of the sponge (see Figure 2a). The inclusion of local defects (simulating wounds and scars of the organism in the hollow cylindrical lattice with helical ridges, see Methods and Extended Data Figure 4 for details) plays a secondary role in the flow velocity at *P*, whereby it contributes modest fluctuations of the order of 10% of the peak-to-peak oscillations in the velocity components experienced by the solid models. The effect of Re on the uniformity of the flow field downstream of the models under investigation is summarized in Figure 2b, where the flow pattern exhibits a symmetry breaking above a critical value of Re. Symmetry breaking for the two solid models occurs at lower critical Re, indicating that the presence of the fenestrae provides a potent stabilizing effect on the flow wake at high Re numbers, downstream of the organism.

To further detail the role in the downstream flow played by the skel- etal motifs of *E. aspergillum*, we analysed the helicity $\mathcal{H}$ and enstrophy $|\boldsymbol{\omega}|^2$ fields (see Methods) of the hollow cylindrical lattice with helical ridges and compared them to those of the plain cylinder (P2 versus S1), as reported in Figure 3. Figure 3 shows a quiescent region downstream of the porous model, which is absent for the solid one. The results are representative of the statistical steady state reached at Re = 2,000, but the same findings are observed in all our simulations with Re ≥ 500 (see Extended Data Figure 5 and Supplementary Videos 1–3). The helicity and enstrophy charts in Figure 3b highlight that the quiescent region extends several diameters downstream of the porous model. The extent of this region is robust with respect to the presence of local defects.

One of the implications of the presence of this nearly quiescent region downstream of the porous model is a reduced hydrodynamic load. In turn, this will mitigate the bending stress experienced by the skeletal system, thereby further contributing to its exceptional mechanical stability that has been so far attributed solely to its mechanical properties[4]. To quantify the hydrodynamic loading experienced by the models, we computed the drag coefficient $C_D$:

$$C_D = \frac{2\bar{\bar{F}}_{drag}}{A\rho_{inlet}u_{inlet}^2}$$

(1)

where $\bar{\bar{F}}_{drag}$ is the total (average) drag force acting on the model, computed at the statistical steady state and along the flow direction; $\rho_{inlet}$ and $u_{inlet}$ are the fluid density and speed at the domain inlet, respectively; and *A* is the area of the transverse section of the model, perpendicular to the fluid flow (the value *A* = *DH* for P1 and S1 is modified for P2 and S2 to account for helical ridges). The values of $C_D$ are reported in Figure 3c. Predictably, the presence of the fenestrae on the surface of *E. aspergillum* yields a remarkable drag reduction, evident for Re ≥ 500. On the other hand, the helical ridges introduce a systematic drag increase, which, however, does not mitigate the benefits provided by the fenestrae (see Figure 3c). These results are robust with respect to the presence of local defects, which are responsible for a modest drag reduction from 2.5% to 3.5% (see Extended Data Figure 6). Since drag reduction is often a goal of sessile aquatic organisms to reduce deformation and prevent break- age[31], the presence of the ridges introduces a new question as to whether they exclusively serve structural functions, such as those identified in ref. [3]. We propose that this is not the case, with the helical ridges regulating a trade-off between drag reduction and feeding and reproductive functions.

In contrast to shallow-water sponges that have access to highly nutritious photosynthetic algae, deep-sea sponges feed mainly on non-photosynthetic bacteria that are suspended at low concentrations in the water column[32]. Feeding is particularly challenging in the habitat of deep-sea sponges, where strong currents favour the presence of clay and detritus (making up more than 97% of suspended particles).

*E. aspergillum* accomplishes highly effective, selective filter feeding[11], through low-speed vortical structures within its body cavity, which are evidenced in Figure 1c. These complex swirling patterns favour the distribution of suspended particles throughout the flagellated chambers of the sponge, where nutrients are absorbed and inorganic particles are discarded. Similarly, the reproduction of *E. aspergillum* is likely to be enhanced through the swirling patterns within the body cavity. Sexual reproduction is known to occur in this species, in which free-spawned sperm fertilize retained eggs[12,20]: vortical structures could thus favour the encounter between gametes within the body cavity.

To quantify the extent to which the helical ridges of E. aspergillum contribute to the generation of these low-speed vortical structures within the body cavity, we contrasted the flow fields of the two porous models (P1 versus P2). Such a comparison was carried out in terms of the vorticity magnitude and the Q-criterion, which defines a vortex as a flow region where $Q = 0.5\,(\|\Omega\|^2 - \|S\|^2) > 0$, where $\Omega$ and $S$ are the skew-symmetric and symmetric parts of the velocity gradient tensor[33–35] (see Figure 4 and Methods for further details). According to such a criterion, $Q > 0$ implies that the energy connected with local fluid rotation prevails over dissipative phenomena, such that vortical structures may arise within the flow. Figure 4a provides evidence of a magnifying effect of the helical ridges on the generation of swirling patterns within the body cavity. From the helicity, we further estimated the residence time within the body cavity, as a measure of the time available for suspended nutrients and sperm to dwell within the sponge body cavity. The presence of the helical ridges is responsible for increasing the mean residence time and widening its distribution, thereby increasing the time available for feeding and sexual reproduction.

Overall, our computational results reveal a rich, multifaceted role of the skeletal motifs of *E. aspergillum* on the flow physics within and beyond its body cavity. Skeletal motifs play a critical role in the functional ecology of this species, beyond the previously known benefits to its mechanical properties. More specifically, the presence of the fenestrae delays the intermittency that arises at Re > 100 several diameters downstream from the organism. This provides a reduction in the drag, which mitigates the stress experienced by the sponge and improves its mechanical stability. The helical ridges that decorate the surface of the sponge promote vortical structures within the body cavity that increase the available time for the sponge to feed and sexually reproduce, at the cost of a secondary increase in the drag.

Our results unveil mechanisms of extraordinary adaptation to live in the abyss, laying the foundations of a new class of computational investigations at the intersection between fluid mechanics, organism biology and functional ecology.

**Online content**

Any methods, additional references, Nature Research reporting summaries, source data, extended data, supplementary information, acknowledgements, peer review information; details of author contributions and competing interests; and statements of data and code availability are available at https://doi.org/10.1038/s41586-021-03658-1.

**Methods**

**Skeletal system of *E. aspergillum***

The skeletal system of this organism comprises four main regions: the anchoring bulb, the curved section connecting the bulb to the main body, the main body, and the terminal sieve plate at the apex, called the osculum (links to high-resolution photographs and images of *E. aspergillum* can be found at: https://commons.wikimedia.org/wiki/ File:Euplectella-aspergillum.jpg; https://asknature.org/strategy/ glass-skeleton-is-tough-yet-flexible/; https://www.researchgate. net/figure/Euplectella-Aspergillum_fig1_344450084; http://www. microscopy-uk.org.uk/mag/indexmag.html?http://www.microscopy- uk.org.uk/mag/artfeb12/rh-euplectella4.html; further images are in Extended Data Figure 2, Extended Data Figure 3, Extended Data Figure 7, and Extended Data Figure 8). We realized a digital mock-up of the complete geometry, as reported in Extended Data Figure 7 and Extended Data Figure 8, by accurately reproducing all its main characteristics in SolidWorks and MeshLab. Excluding the solid anchoring bulb, the remaining regions consist of a periodic lattice of two intersecting patterns. The main pat- tern is composed of axial and circumferential filaments with a diameter of 0.5 mm, orthogonally crossing each other. The secondary pattern is characterized by smaller ligaments with a diameter of 0.2 mm and is arranged at 45° with respect to the main pattern.

The lattice envelops a cylinder of 40 mm in diameter that defines the main body section and bends into a cone-like shape that represents the connecting region to the anchoring bulb. A non-periodic arrangement of 5-mm-thick helical ridges is placed on the outer surface of the lattice. Ridges are organized in a sequence of rows along the secondary pattern of the lattice, with intersecting helical patterns and random interruptions according to the literature[3]; simulations on this complete geometry are in Figure 1c.

Along with the detailed reconstruction of *E. aspergillum*'s skeletal system, four simplified models of the main body were generated (see Figure 1b) to assess the effect of morphological manipulations on the flow physics, namely, a plain cylinder (S1), a plain cylinder with helical ridges (S2), a hollow cylindrical lattice (P1) and a hollow cylindrical lattice with helical ridges (P2); simulations on these models are in Figure 2, Figure 3, and Figure 4.

Extended Data Figure 1 demonstrates a lattice site resolution within the fenestrae of the secondary ligament pattern of the hollow cylindrical lattice with helical ridges (and its manipulations) of ~5 × 5 lattice sites, which is sufficient for the resolution of the hydrodynamics with the lattice Boltzmann method[42]. More specifically, for the method to be reliable, the mean free path $\lambda$ should be much smaller than the lattice spacing, so that $\lambda = c_s (\tau - 0.5) \ll 1$, with $c_s$ being the lattice sound speed and $\tau$ the characteristic timescale towards local relaxation. For Re = 2,000, we have $\nu = 0.01$ lattice units, corresponding to $\tau = 0.5033$, which implies that in the ~5 grid points of the side of the fenestra, fluid particles collide $O(100)$ times, thereby ensuring the hydrodynamic regime through the hole.

Beyond the validation of our predictions on drag coefficients at Re = 100 in Figure 3c, we garnered further evidence through the analysis of the wake downstream of the plain cylinder model (S1). More specifically, we examined the Strouhal number, $St = \mathcal{F} D / u_{inlet}$, where $\mathcal{F}$ is the frequency of the downstream vortex shedding (evaluated at the probe *P* location), *D* is the model characteristic length (that is, the diameter), and $u_{inlet}$ the inlet velocity magnitude. Extended Data Table 3 summarizes the comparison between our predictions of St at different Re against experimental measurements[43]. Extended Data Table 3 demonstrates the accuracy of our numerical predictions, with an error ε ranging between 3.8% and 6.3%. Standard Triangulation Language (STL) files of *E. aspergillum* and periodic models are available as Supplementary Data.

**Random defects in the geometry**

Taking advantage of the versatility of our numerical method to explore models of arbitrary complexity, we generated nine variations of the hollow cylindrical lattice with helical ridges (see Figure 1b). These variations were obtained by applying random morphological defects in the skeletal motifs that simulate wounds and scars, which have been widely documented[20]. Wounds were simulated by removing filaments and portions of ridges in the pristine model; conversely, scars were represented as added masses to the lattice and the ridges (see Extended Data Figure 4).

The mass of the model was kept constant across variations, which accounted for a ~5% mass redistribution. The rationale for this selection is grounded in the literature[20], which indicates that these organisms can regenerate no more than 10% of their skeletal system when dam- aged, while exceeding such a threshold would lead to death. We report the nine variations of the hollow cylindrical lattice with helical ridges (P2) that were obtained by applying random morphological defects in Extended Data Figure 4. We performed extreme fluid dynamic simulations for these nine variations, covering all of the considered Re (namely, Re = 100, 500, 1,000, 1500 and 2,000). The data set was used to ascertain statistical variations on the drag coefficient $C_D$, the non-dimensional residence time $t^*$, and helicity $\mathcal{H}$ and enstrophy $|\omega|^2$ within the computational domain.

**The lattice Boltzmann method**

The fluid fields evolve in time according to the following lattice Boltzmann equation:

$$f_i(\mathbf{x} + \mathbf{c}_i, t + 1) - f_i(\mathbf{x}, t) = \frac{1}{\tau}[f_i^{eq}(\mathbf{x}, t) - f_i(\mathbf{x}, t)]$$

Where $f_i(\mathbf{x}, t)$ represents the probability density function of finding a fluid particle at site **x** and discrete time $t$, moving along the $i$th lattice direction ($\mathbf{c}_i$), and $f_i^{eq}$ identifies local Maxwellian equilibrium. In the present work, we employed a 19 discrete speed ($i = 0,...,18$) scheme along the three spatial dimensions, also known as a D3Q19 lattice. For details on the method, we refer to refs. [7,44–46].

One of the main advantages of the lattice Boltzmann method is the handling of complex geometries, such as the one of *E. aspergillum* under investigation. Taking advantage of the discrete Cartesian nature of the method, it is possible to implement second-order accurate boundary conditions by using the algorithms proposed in refs. [7,45]. Here, we impose the following boundary conditions:

•Inflow: we fix the velocity magnitude based on the desired Re (ref. [45]);

•Outflow: we impose a zero-gradient condition, by copying the density and velocity from the last fluid plane into the boundary outlet buffer[45];

•Side boundaries: we use periodic boundary conditions;

•Internal boundaries: we impose no-slip boundary conditions via the standard bounce-back procedure. This condition is applied to the solid nodes of the geometries (S1, S2, P1, P2 and the complete model of *E. aspergillum*). The same no-slip boundary conditions are also employed for the seafloor in the study of the complete model; and

•Top-bottom boundaries: We use periodic boundary conditions for all geometries, except of the complete model, for which we enforce no-slip on the bottom and free-slip on the top.

To examine the flow physics inside the body cavity and in the down- stream wake of the considered geometries, we studied the flow vorticity **ω**, enstrophy $|\omega|^2$, and helicity $\mathcal{H}$. Through these quantities, we identified the presence of vortical structures via the *Q*-criterion[34], and we estimated the non-dimensional residence time within the body cavity; more details on these computations can be found elsewhere[47]. For the sake of completeness, we report below the definitions of the various fluid parameters employed.

Enstrophy is given by

$$|\boldsymbol{\omega}|^2 = |\nabla \times \mathbf{u}|^2 = \left(\frac{\partial u_z}{\partial y} - \frac{\partial u_y}{\partial z}\right)^2 + \left(\frac{\partial u_x}{\partial z} - \frac{\partial u_z}{\partial x}\right)^2 + \left(\frac{\partial u_y}{\partial x} - \frac{\partial u_x}{\partial y}\right)^2$$

whereas helicity is computed as

$$\mathcal{H} = \mathbf{u} \cdot (\nabla \times \mathbf{u}) = u_x \left(\frac{\partial u_z}{\partial y} - \frac{\partial u_y}{\partial z}\right)^2 + u_y \left(\frac{\partial u_x}{\partial z} - \frac{\partial u_z}{\partial x}\right)^2 + u_z \left(\frac{\partial u_y}{\partial x} - \frac{\partial u_x}{\partial y}\right)^2$$

Helicity in the body cavity is in turn used to estimate the non-dimensional residence time.

To compute the Q-criterion, we separated the antisymmetric (vorticity tensor) and symmetric (rate-of-strain tensor) components of the velocity gradient tensor, defined as follows:

$$\bar{\Omega} = \frac{1}{2}[\nabla \mathbf{u} - (\nabla \mathbf{u})^T] = \frac{1}{2}\left(\frac{\partial u_\alpha}{\partial \beta} - \frac{\partial u_\beta}{\partial \alpha}\right)$$

$$\bar{S} = \frac{1}{2}[\nabla \mathbf{u} + (\nabla \mathbf{u})^T] = \frac{1}{2}\left(\frac{\partial u_\alpha}{\partial \beta} + \frac{\partial u_\beta}{\partial \alpha}\right)$$

where T indicates tensor transposition and Greek subscripts $\alpha$, $\beta$ label the Cartesian components. Finally, the Q factor is defined as

$$Q = \frac{1}{2}[\|\bar{\Omega}\|^2 - \|\bar{S}\|^2]$$

where the symbol $\|\cdot\|$ denotes tensorial norm. Positive values of Q identify flow regions where rotational energy exceeds dissipation, thereby setting the stage for the emergence of vortical structures. High helicity, on the other hand, associates with the presence of three-dimensional spiralling structures.

Extended Data Table 2 reports the main physical and computational parameters, including information about the conversion between lattice and physical units.

**High-performance computing**

Simulations were carried out on two supercomputing facilities at CINECA, namely, Marconi and Marconi100. The former is a CPU based architecture, while the latter, ranking ninth in the May 2020 Top500 list[17], is based on GPU accelerators. Simulations on S1, S2 and P1 were performed on Marconi, involving 128 tasks and 64 threads for each task, for a total of 8,192 cores used. These are the least expensive runs among those we performed, yielding 0.7 petaflops (peta floating point operations per second) peak performance and exploiting ~4% of the CPU-based high-performance computing facility. Simulations on P2 and its nine variations with random defects were performed on Marconi100. Each run used 16 compute nodes: each compute node has 32 IBM Power9 cores and 4 Nvidia V100 GPUs (with 80 stream- ing multiprocessors per GPU). All of the simulations on the complete model of *E. aspergillum* were run on 128 compute nodes of Marconi100, involving one-eighth of the entire high-performance computing facil- ity and yielding ~4 petaflops of peak performance. The simulations on the complete model required 4,096 CPU cores and 40,960 streaming multiprocessors, on a domain characterized by $O(100 \times 10^9)$ grid points for ~$5 \times 10^6$ time steps.

The overall computational effort for all of the simulations presented in this study resulted in $O(10^2)$ terabytes of raw data, and it required ~75,000 GPU hours and ~2,000,000 CPU hours. Extended Data Table 3 summarizes the architecture and computational resources of the study. The excellent scalability of the lattice

Boltzmann method on the aforementioned computational architecture proved instrumental in enabling the full-scale simulation of *E. aspergillum*, from microscopic geometric details, all the way up to the entire organism.

**Supporting fluid dynamics analyses**

Simulation results on vorticity magnitude for all Re are presented in Extended Data Figure 5 to verify the formation of a nearly quiescent region downstream the porous models at Re ≥ 500. A zoomed-out view of Figure 3c with error bars associated with the minimum and maximum values is shown in Extended Data Figure 6 to assess statistical variations in the prediction of the drag coefficient due to the random defects.

Along with the drag coefficients, we also considered the lift coefficient, defined as

$$C_L = \frac{2 F_{lift}}{A \rho_{inlet} u_{inlet}^2}$$

where $F_{lift}$ is the time-varying in-plane transverse force experienced by the models. For Re = 2,000, we report in Extended Data Figure 9, the time trace at statistical steady state of the lift coefficient.

**Data availability**

STL files for all of the models, raw data for the plots, and scripts to reproduce the figures are available on GitHub at https://github.com/giacomo falcucci/Euplectella_HPC. Additional data that support the findings of this study are available from the corresponding author on request.

**Code availability**

All codes necessary to reproduce results in main paper are available on GitHub at https://github.com/giacomofalcucci/Euplectella_HPC.

**Acknowledgements**

G.F. acknowledges CINECA computational grant ISCRA-B IsB17– SPONGES, no. HP10B9ZOKQ and, partially, the support of PRIN projects CUP E82F16003010006 (principal investigator, G.F. for the Tor Vergata Research Unit) and CUP E84I19001020006 (principal investigator, G. Bella). G.P. acknowledges the support of the Forrest Research Foundation, under a postdoctoral research fellowship. M.P. acknowledges the support of the National Science Foundation under grant no. CMMI 1901697. S.S. acknowledges financial support from the European Research Council under the Horizon 2020 Programme advanced grant agreement no. 739964 ('COPMAT'). G.F. and S.S. acknowledge K. Bertoldi, M. C. Fernandes and J. C. Weaver (Harvard University) for introducing them to *E. aspergillum* and for early discussions on the subject. A. L. Facci (Tuscia University) is acknowledged for discussions on graphics realization. M. Bernaschi (IAC-CNR) is acknowledged for discussions on extreme computing. V. Villani is acknowledged for his support with Japanese language and culture. E. Kaxiras (Harvard University) is acknowledged for early discussions that proved fruitful for the development of the present code.

**Author contributions**

G.F. designed the research; G.F. and G.A. wrote the original lattice Boltzmann method code; G.A. extended the code for massively parallel computation, developed the GPU version for Marconi100, and helped collect and post-process the data;

P.F. realized all of the models; V.K.K. ran the validation tests and helped in post-processing and data interpretation; G.F. created the figures; G.P. and M.P. led the biological framing of the results; G.F., M.P. and S.S. supervised the research and the interpretation of the results; G.F., G.P., M.P. and S.S. wrote the manuscript. All authors contributed to analysing the results of the simulations and revising the manuscript.

**Figures**

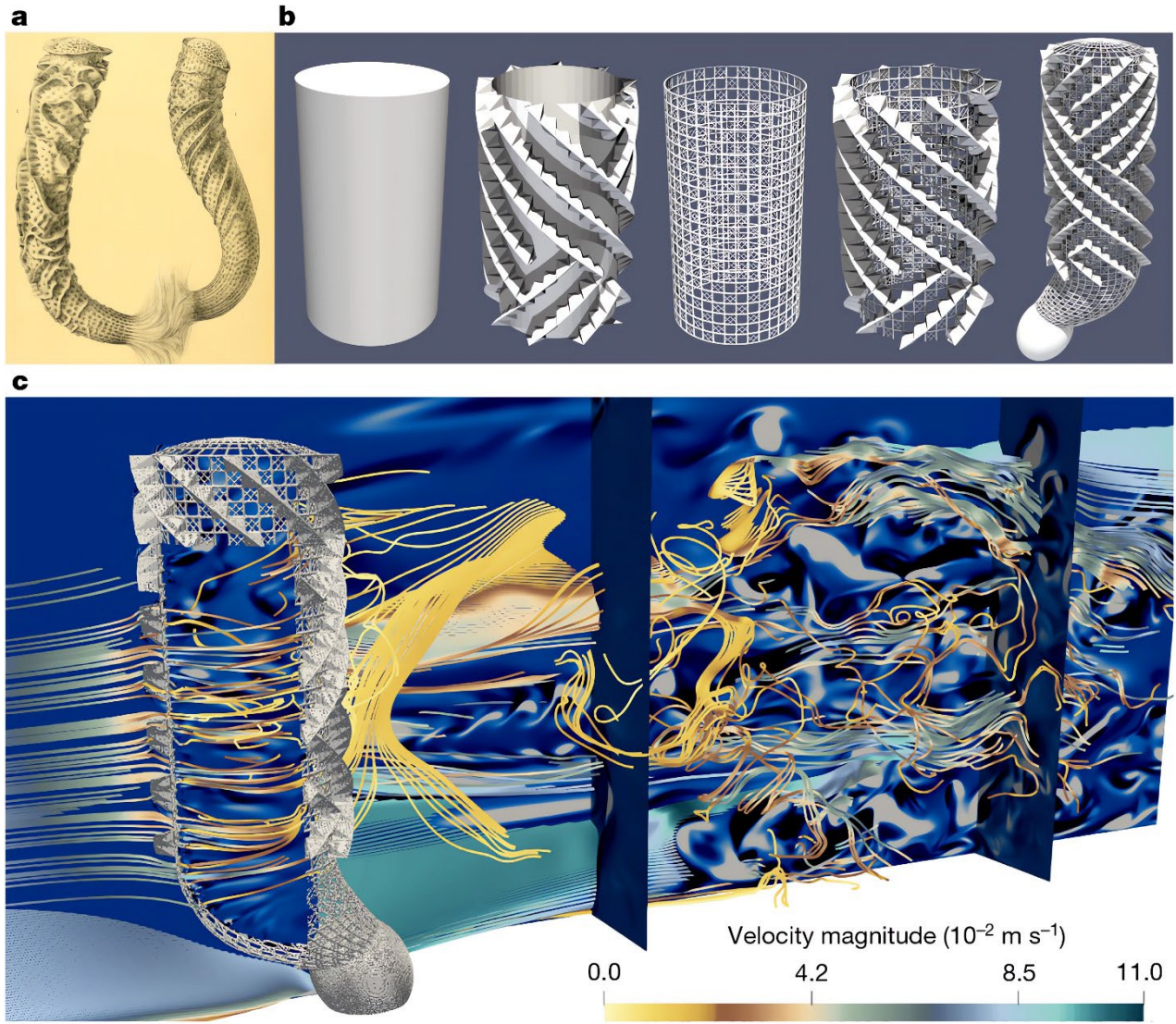

*Figure 1: Skeletal motifs of E. aspergillum and associated flow physics.* **a**, Original drawing[2] *of the deep-sea glass sponge E. aspergillum.* **b**, *Models used in this work, showing geometric progression of increasing complexity towards the structure of E. aspergillum. Left to right: solid model S1 (plain cylinder), solid model S2 (cylinder with helical ridges), porous model P1 (hollow cylindrical lattice), porous model P2 (hollow cylindrical lattice with helical ridges), and a complete model of E. aspergillum, reconstructed according to ref.* [3]. *All models are generated with a spatial accuracy of 0.2 mm (see Methods and Extended Data Fig. 1).* **c**, *Simulation showing the complete model of E. aspergillum immersed in a hydrodynamic flow at Re = 2,000. The panel shows contours of the helicity and streaklines of the flow, coloured according to the flow velocity magnitude (colour scale at bottom right). Extreme simulations capture the formation of the boundary layer on the seafloor and its interaction with the skeletal motifs of E. aspergillum. Low-speed vortical structures within the sponge arguably favour selective filter feeding and gamete encounter for sexual reproduction. The region of near quiescence extends downstream of the sponge, moderating the hydrodynamic loading experienced by the organism, as highlighted by the two vertical cross-sections of the downstream helicity field.*

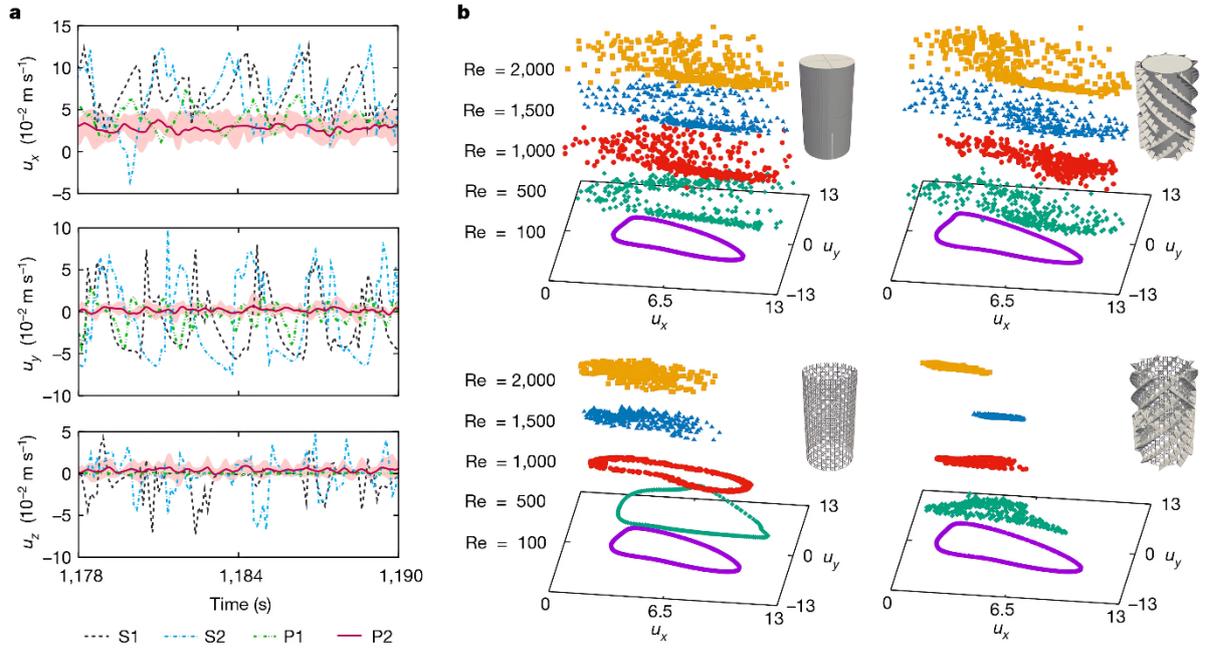

*Figure 2: Effect of manipulations of the morphology of E. aspergillum on the flow downstream. **a**, Simulated time evolution of the three components of the flow velocity **u**(P, t), where P is a probe located 2.5 diameters downstream of the model, at Re = 2,000, for the four considered, periodic geometries (models S1, S2, P1, P2; see key). The comparison of the velocity components (top to bottom, panels show $u_x$, $u_y$, $u_z$) at P for the different models confirms the abating effect of the skeletal motifs of E. aspergillum on flow fluctuations downstream of the sponge. The panels report the last ~1% of the whole simulation time span (5 × 10$^4$ out of 5.2 × 10$^6$ time steps, corresponding to the last ~20 s of the simulated time). Data include statistical variations due to local defects (solid lines are mean values, and shaded regions identify minima and maxima). **b**, For each model (top left, S1; top right, S2; bottom left, P1; bottom right, P2) we show polar diagrams of the x, y velocity components in cm s$^{-1}$, downstream of the model, at probe location P, for all explored Re regimes (data are colour coded to indicate Re value, given at left). The panel highlights the stabilizing effects on the fluid wake due to the concurrent influence of the fenestrae and the ridges of E. aspergillum.*

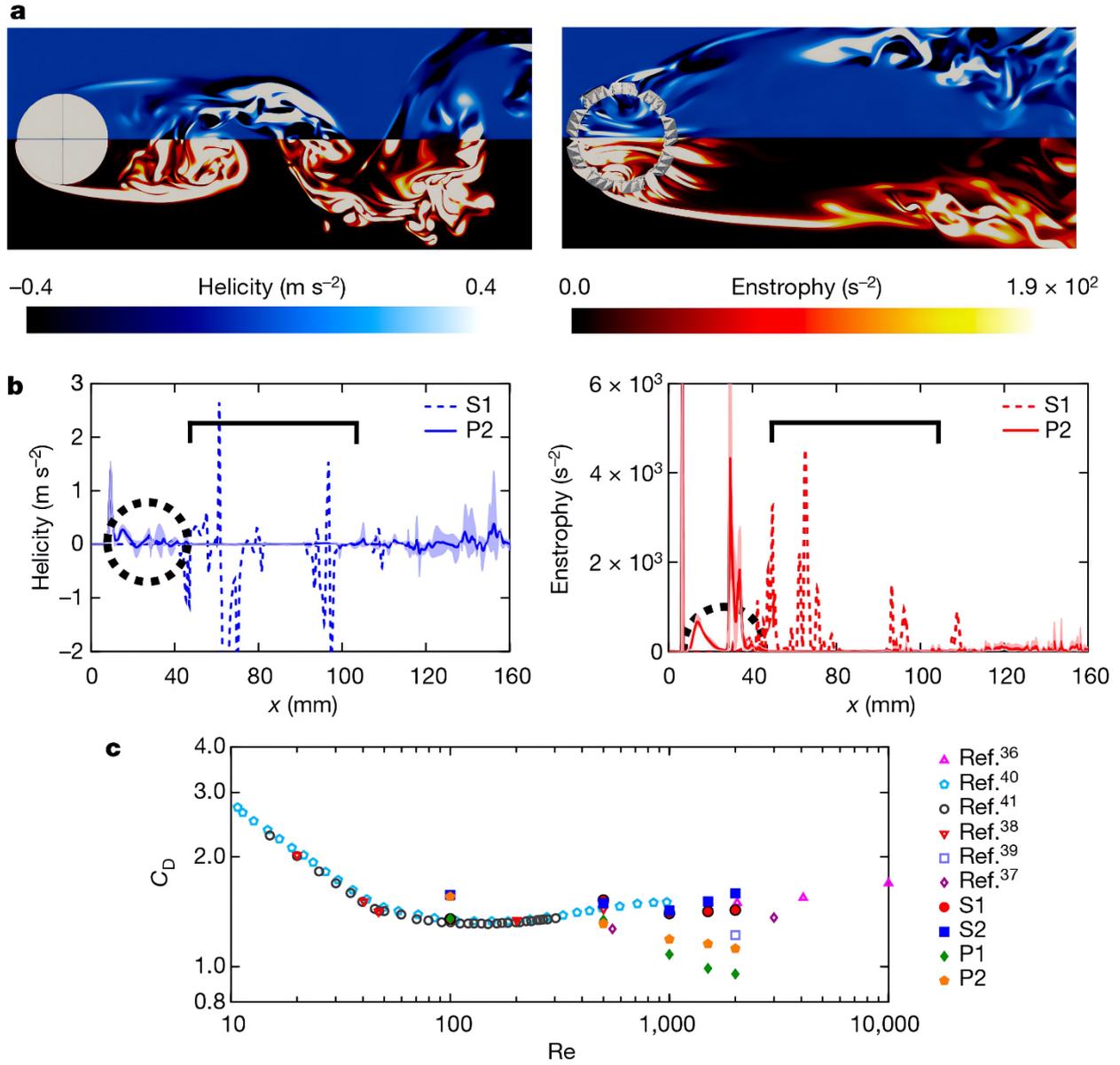

Figure 3: Effect of manipulations of the morphology of E. aspergillum on helicity, enstrophy and drag coefficient. **a**, Helicity $\mathcal{H}$ (upper part of each panel) and enstrophy $|\boldsymbol{\omega}|^2$ (lower part of each panel) fields at Re = 2,000 for the plain cylinder (S1, left panel) and the hollow cylindrical lattice with helical ridges (P2, right panel). Colour scales for helicity and enstrophy are shown under. **b**, Zoomed-in view of helicity (left panel) and enstrophy (right panel) along the x direction in the centreline of the domain for S1 and P2 (see keys) at Re = 2,000; data include statistical variations due to local defects (solid lines are mean values, and shaded regions identify minima and maxima). The dashed circle identifies the model and the black markers the nearly quiescent region that forms downstream of P2, due to its fenestrae and the external ridges. **c**, Drag coefficient $C_D$ for all of the simplified models, compared to literature values for cylinders[36–41]; the comparison at Re = 100 supports the accuracy of the simulations (further validation can be garnered from the Strouhal number; see Methods and Extended Data Table 1). The panel highlights the beneficial role of the fenestrae and the detrimental role of the helical ridges on the drag experienced by the models: the hollow cylindrical lattice with helical ridges offers the second smallest drag coefficient, after the hollow cylindrical lattice. Notably, the helical ridges contribute a reduction of in-plane transverse force, but peak-to-peak oscillations are secondary with respect to the drag force (see Methods and Extended Data Fig. 9).

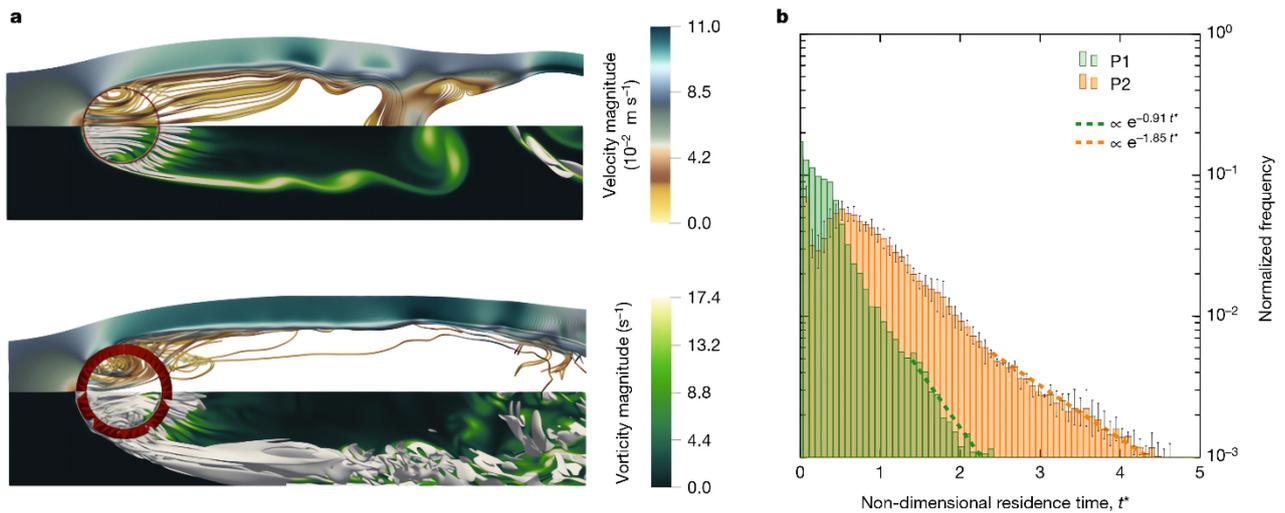

Figure 4: Role of the ridges in flow speed, vorticity, Q-structures and residence time within the body cavity. **a**, Flow speed, vorticity and Q-structures for the two porous models, without and with helical ridges (P1, top panel; P2, bottom panel). The upper part of each panel reports the streaklines coloured according to the flow speed (velocity magnitude, upper colour scale), while the lower part shows contours of the vorticity magnitude |**ω**| (lower colour scale) along with regions characterized by Q > 0 (vortical structures). The panels help to visualize the hydrodynamic role of the external ridges in amplifying vortical structures within the E. aspergillum's body cavity, promoting selective filter feeding and gamete encounter for sexual reproduction. **b**, Distribution of the non-dimensional residence time within the body cavity, $t^* = \mathcal{H}D/u_{inlet}^2$ ; data include statistical variations due to local defects (solid lines are mean values and error bars identify minima and maxima). 'Normalized frequency' refers to the number of occurrences divided by the total number of readings. The panel highlights the effect of the external ridges in extending the tails of the distribution, thereby granting more time to the organism to feed and sexually reproduce. Dashed lines refer to the linear regression of the tail of the distributions.

**Extended Data**

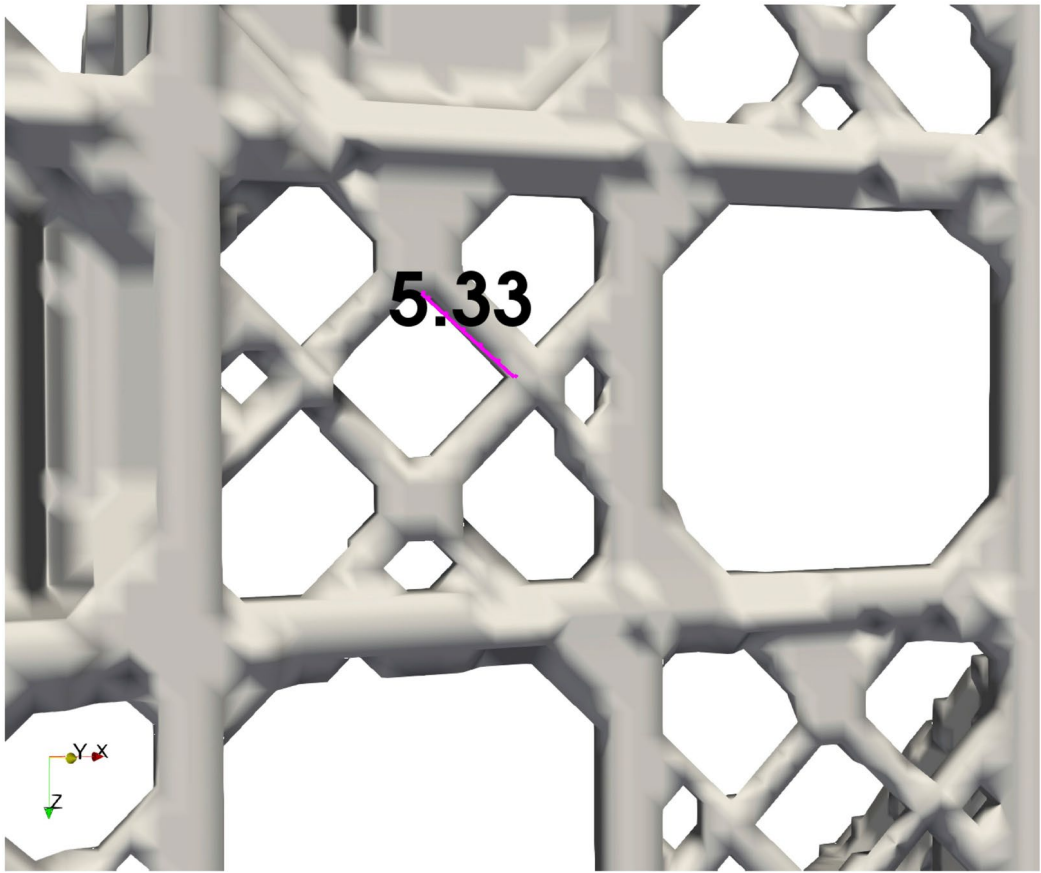

Extended Data Figure 1: Detail of the grid resolution. The grid resolution within the small fenestrae of E. aspergillum models is 5.33 lattice spacings.

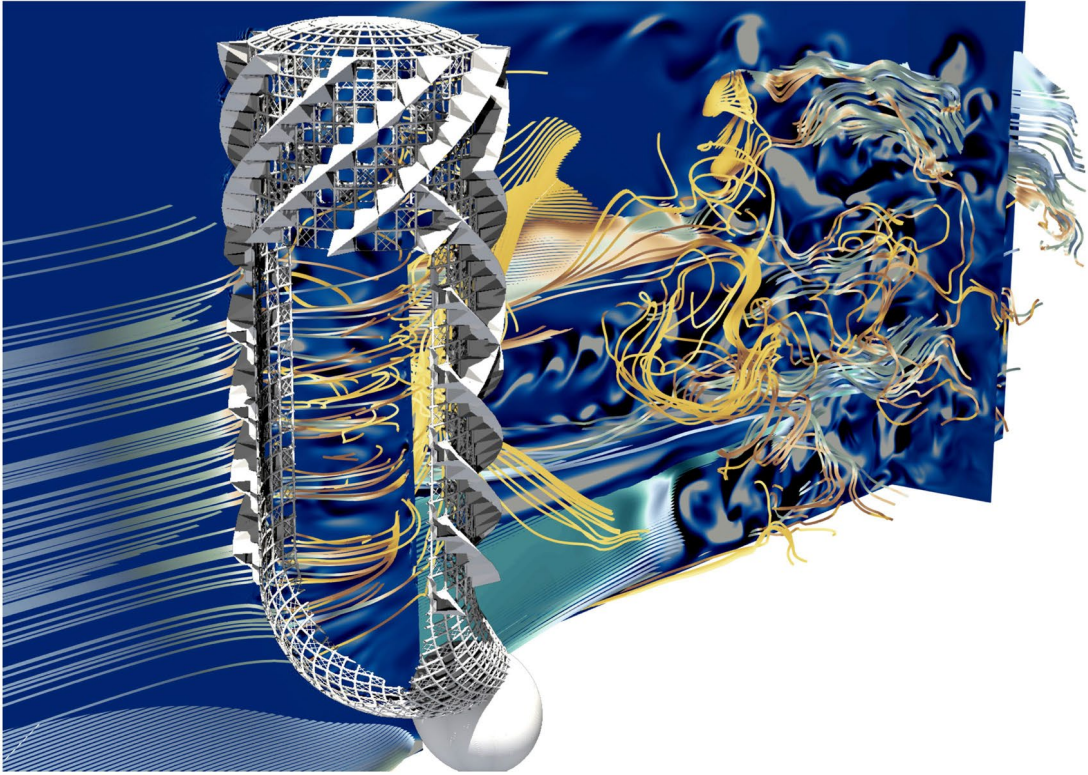

(a)

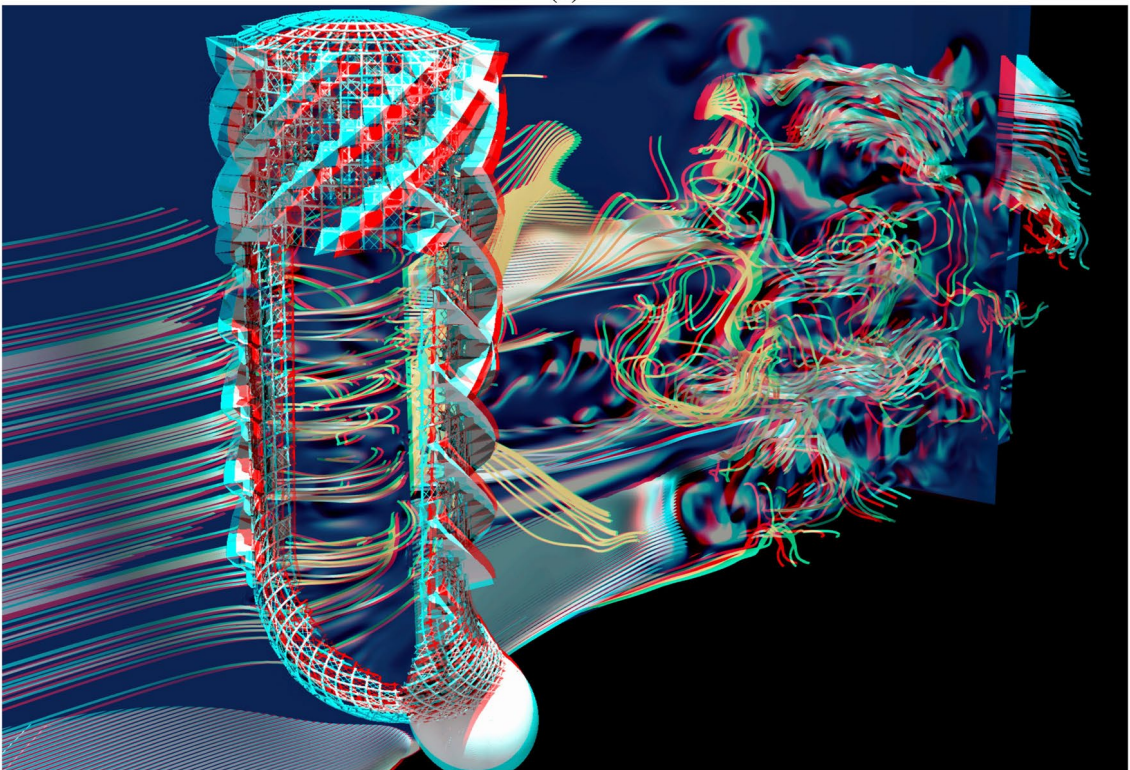

(b)

*Extended Data Figure 2: Details of the flow field. **a**, Tilted view of main text Fig. 1c, detailing the flow field downstream and within the body cavity of the complete model of E. aspergillum at Re = 2,000. Colour intensity indicates the helicity magnitude, and the streak lines are coloured according to the velocity magnitude. **b**, Stereo view of **a**.*

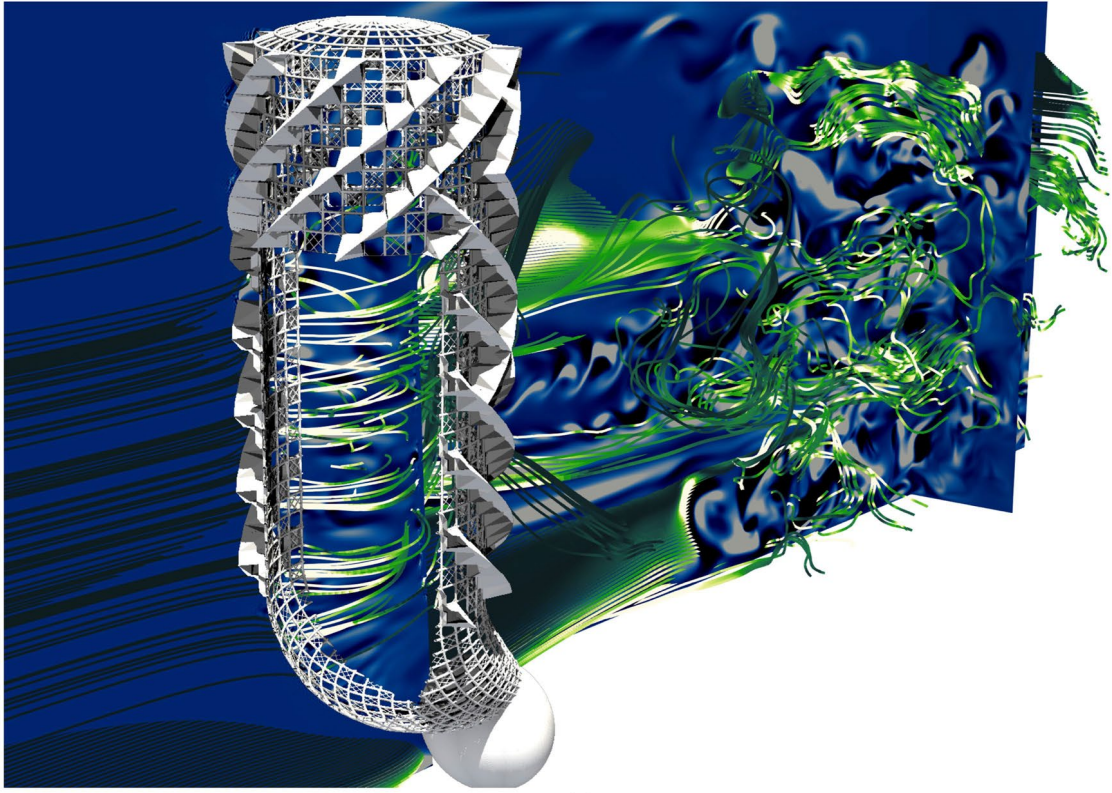

(a)

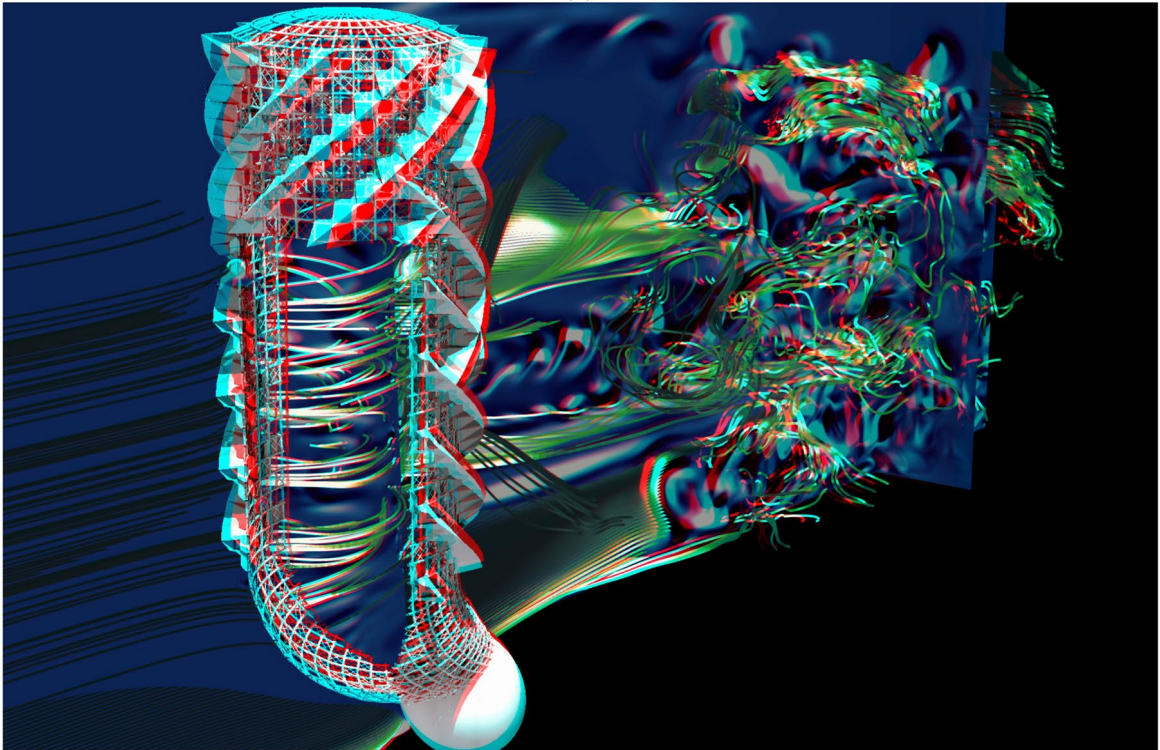

(b)

*Extended Data Figure 3: Details of the vorticity field. **a**, Visualization of the vorticity magnitude, complementing Extended Data Fig. 2a, such that colour intensity indicates the helicity magnitude, and the streak lines are coloured in green, based on the vorticity magnitude. **b**, Stereo view of **a**.*

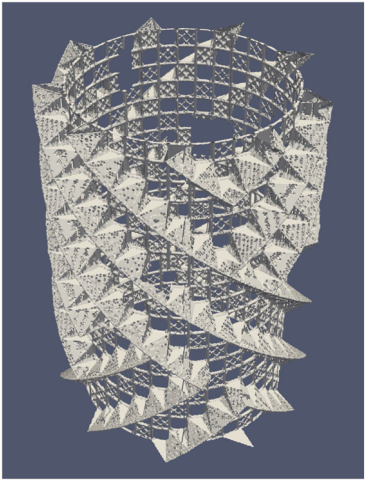 (a) Mark01
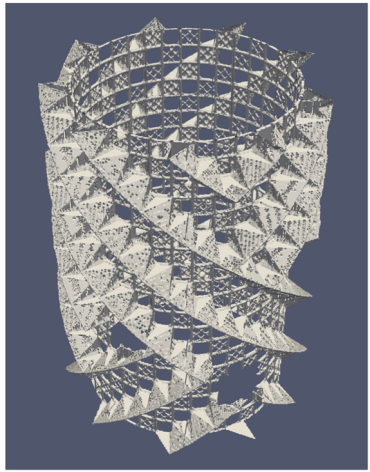 (b) Mark02
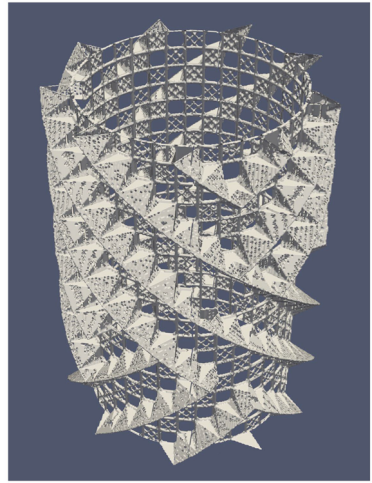 (c) Mark03
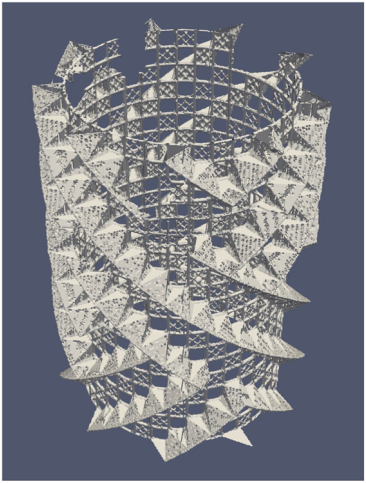 (d) Mark04
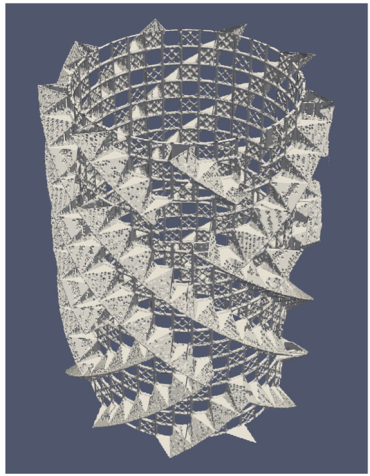 (e) Mark05
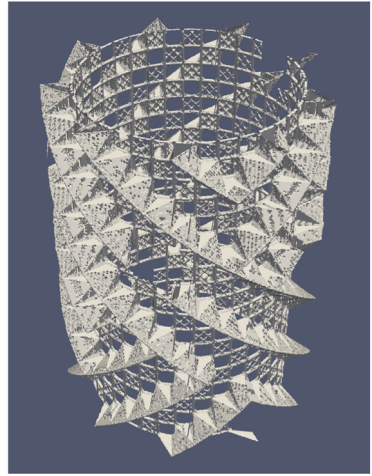 (f) Mark06
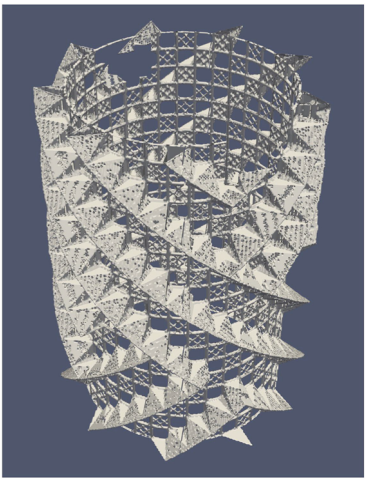 (g) Mark07
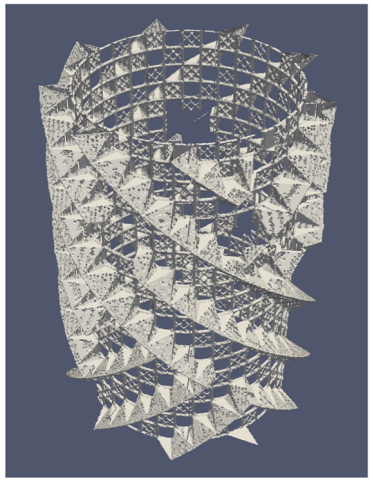 (h) Mark08
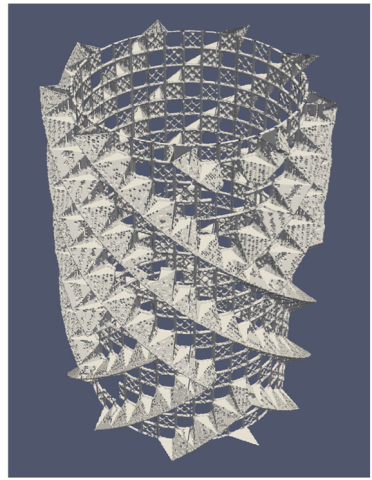 (i) Mark09

*Extended Data Figure 4: Morphological manipulations of the E. aspergillum model. **a–i**, Details of the nine variations of the hollow cylindrical lattice with helical ridges (P2), obtained by including random defects simulating wounds and scars. The nine morphological manipulations are identified as Mark01, Mark02, ..., Mark09.*

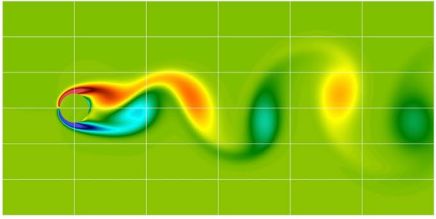 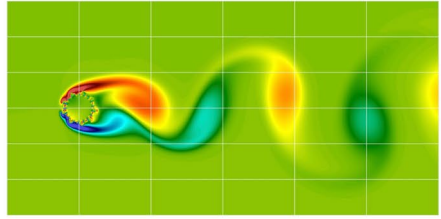

(a) Re = 100

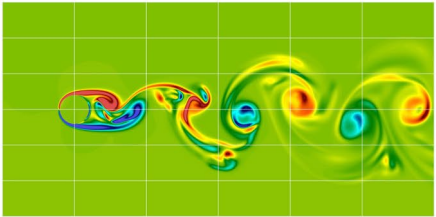 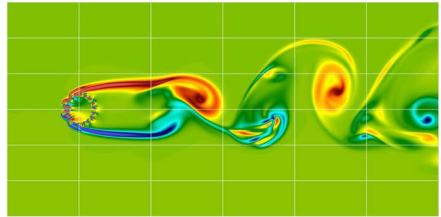

(b) Re = 500

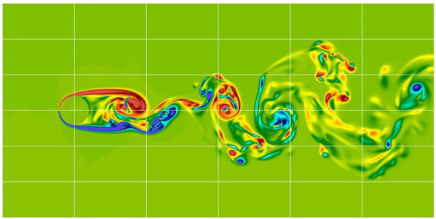 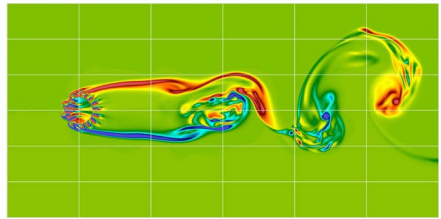

(c) Re = 1000

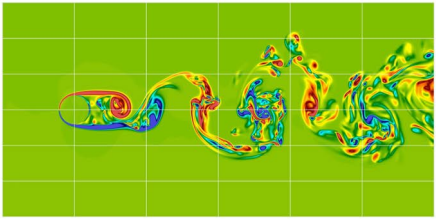 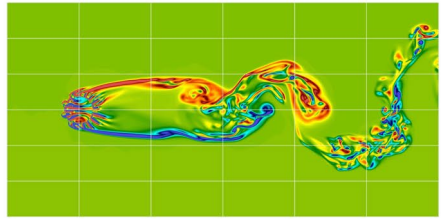

(d) Re = 1500

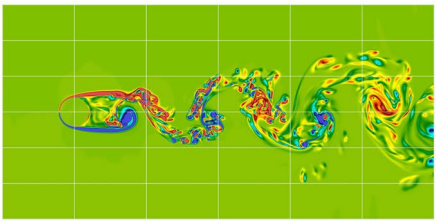 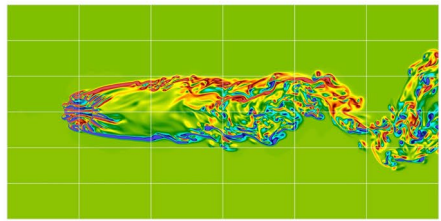

(e) Re = 2000

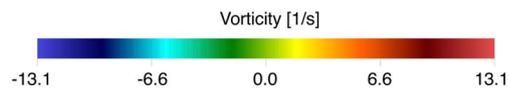

*Extended Data Figure 5: Details of the vorticity magnitude fields. **a–e**, Comparison between the vorticity magnitudes (colour scale) for the plain cylinder (S1, left panels) and for the hollow cylindrical lattice with helical ridges (P2, right panels) at statistical steady states, for all Re simulated in the present work. Panels **a–e** show data for Re = 100, 500, 1,000, 1,500 and 2,000, respectively.*

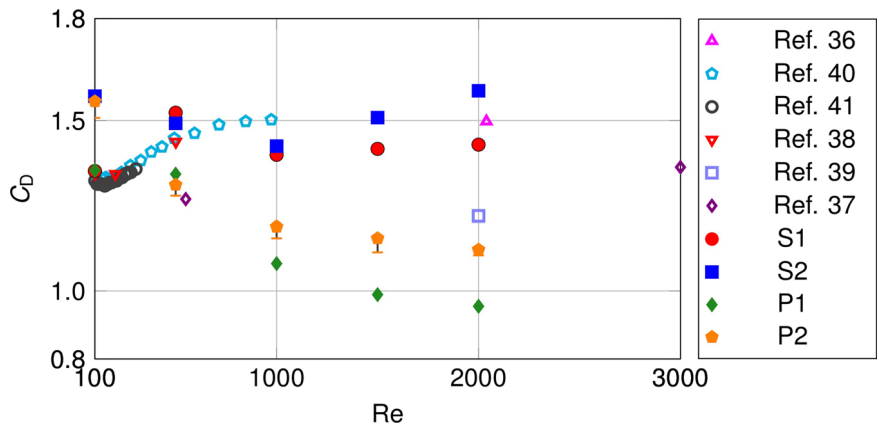

*Extended Data Figure 6: Details of the drag coefficient. Zoomed-out view of main text Fig. 3c, with error bars identifying the range of predicted values of the drag coefficient $C_D$ due to random morphological manipulations. These variations lead to a modest decrease, from 2.5% to 3.5% in the drag coefficient with respect to the pristine model.*

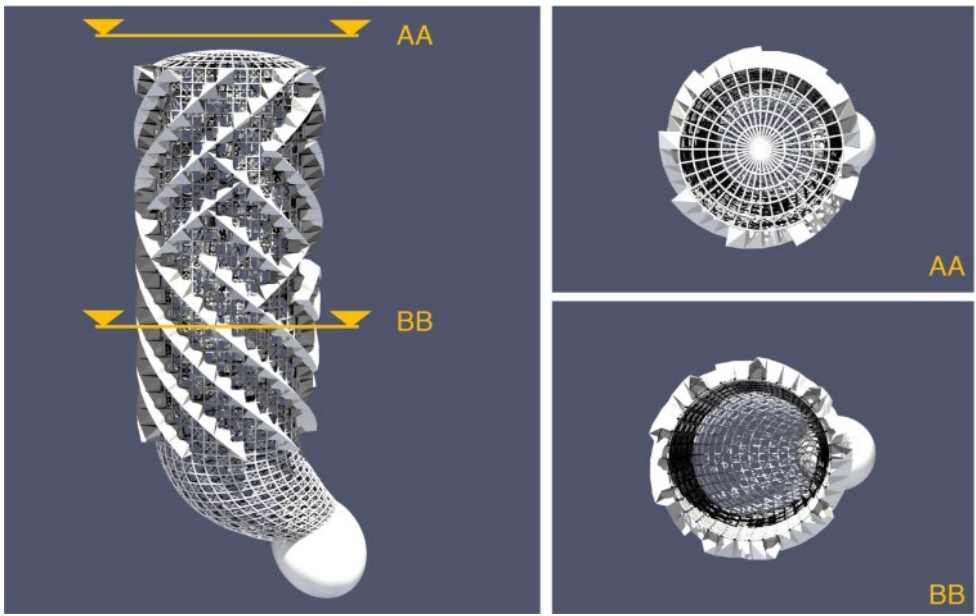

*Extended Data Figure 7: Views of the skeletal system of E. aspergillum. The model is reconstructed according to ref. 3: left, side view; right, AA and BB cross-sections from the left panel, detailing the osculum and the body cavity, respectively.*

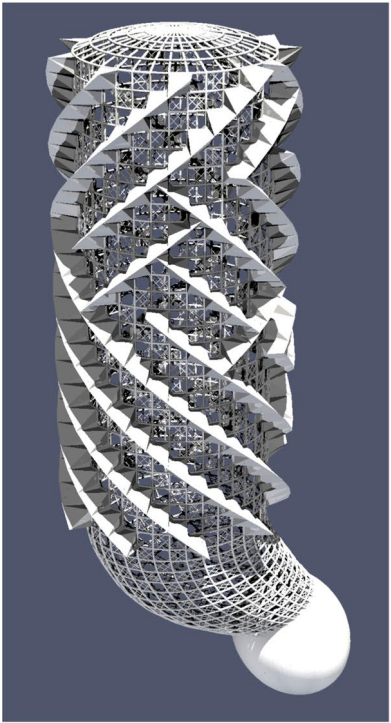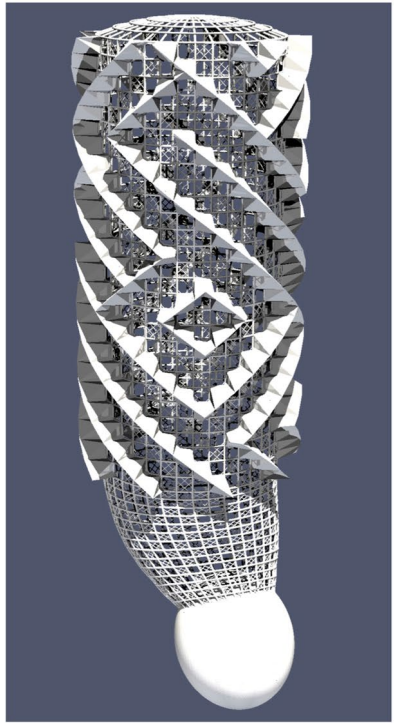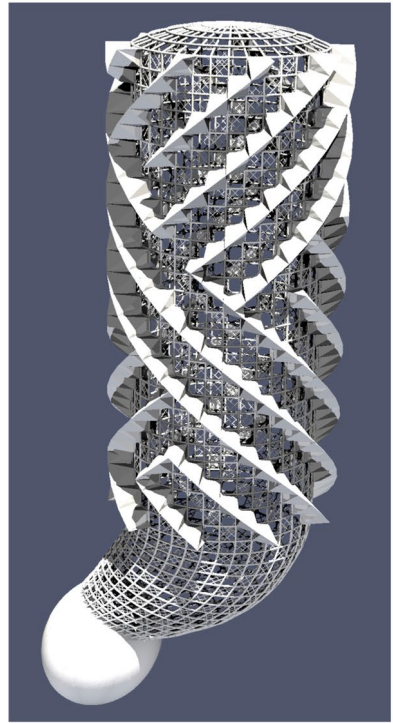

(a)

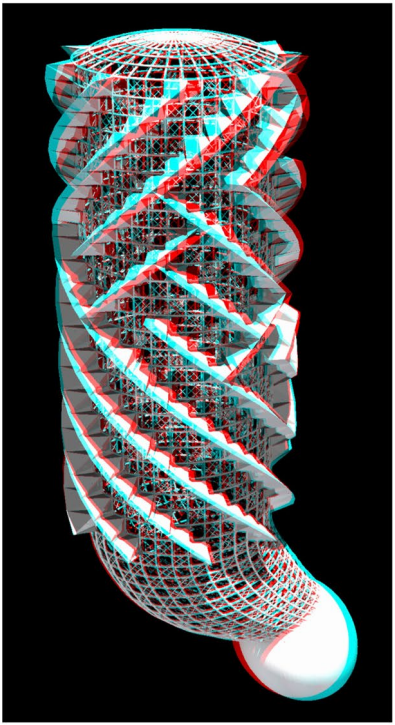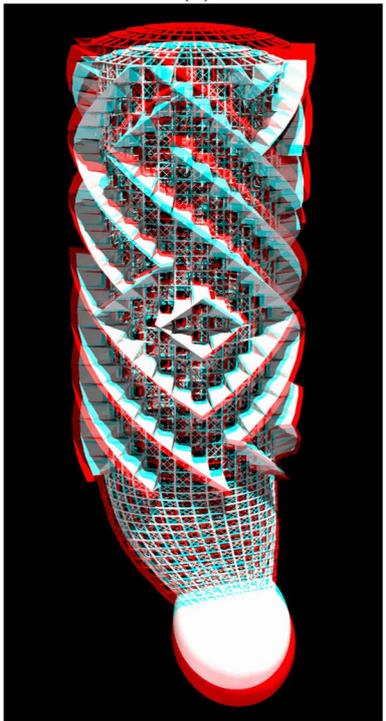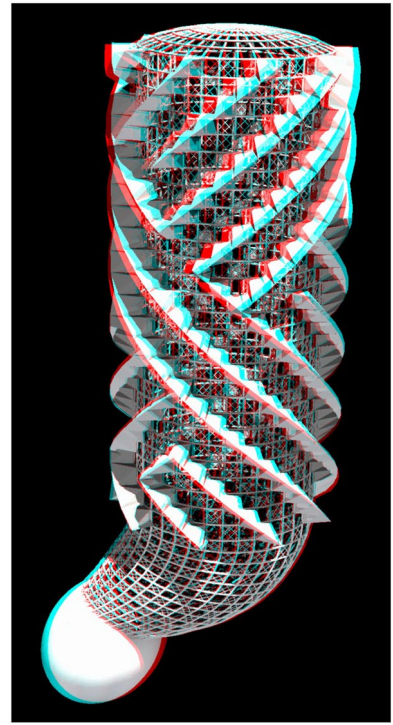

(b)

*Extended Data Figure 8: Details of the E. aspergillum complete model. **a**, Front (centre panel) and side (leftmost and rightmost) views of the complete model of E. aspergillum; **b**, stereo views of the complete model of E. aspergillum realized with the Anaglyph algorithm.*

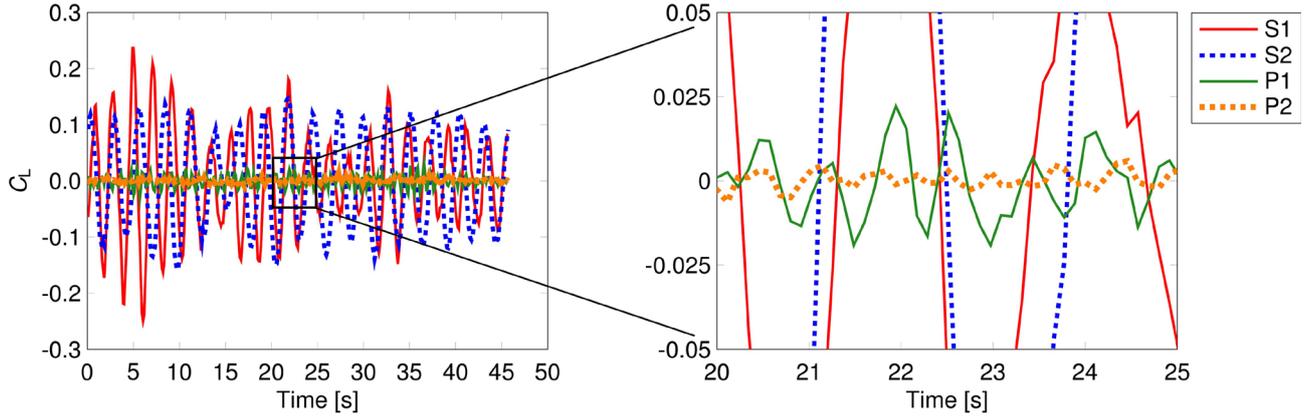

Extended Data Figure 9: Lift coefficient $C_L$. Left, time trace at statistical steady state of the lift coefficient $C_L$ for the different models (see key at right) of E. aspergillum at Re = 2,000. The range of the oscillations in the porous models is two orders of magnitude less than that of the plain cylinder S1. Right, magnified view of boxed region.

| Re | $St_{LB}$ | $St_{exp}$ | $\varepsilon$ |
|---|---|---|---|
| 500 | 0.219 | 0.206 | 6.3% |
| 1000 | 0.223 | 0.210 | 6.2% |
| 1500 | 0.219 | 0.211 | 3.8% |
| 2000 | 0.219 | 0.211 | 3.8% |

Extended Data Table 1: Accuracy assessment. Comparison between numerical and experimental values of St ($St_{LB}$ and $St_{exp}$, respectively), at different Re: $\varepsilon = (St_{LB} - St_{exp})/St_{exp}$.

| Parameters (Re = 2000) | Lattice Units | Physical Units |
|---|---|---|
| Diameter | 200 | 0.04 [m] |
| Flow Speed (@ Inlet) | 0.1 | 0.0875 [m/s] |
| Water Viscosity | 0.01 | $1.75 \cdot 10^{-6}$ [m$^2$/s] |
| Time | 1 | $2.29 \cdot 10^{-4}$ [s] |

Extended Data Table 2: Main physical parameters at Re = 2,000. From the combination of these fundamental quantities, all of the conversion scales for the parameters of interest can be retrieved.

| Geometry | Re | Architecture | # CPU Cores | # GPU |
|---|---|---|---|---|
| S1 | 100, 500, 1000, 1500, 2000 | "Marconi" | 8,192 | - |
| S2 | 100, 500, 1000, 1500, 2000 | "Marconi" | 8,192 | - |
| P1 | 100, 500, 1000, 1500, 2000 | "Marconi" | 8,192 | - |
| P2 | 100, 500, 1000, 1500, 2000 | "Marconi100" | 2,048 | 64 |
| P2 - 9 variations | 100, 500, 1000, 1500, 2000 | "Marconi100" | 2,048 | 64 |
| Complete model of E. aspergillum | 100, 500, 1000, 1500, 2000 | "Marconi100" | 4,096 | 512 |

Extended Data Table 3: Details of resource allocation. For each model (leftmost column), we show the computer resource allocation for the Re numbers examined.